\documentclass[aps,prl,showpacs,twocolumn,superscriptaddress]{revtex4-1}

\usepackage{color,graphicx} 

\begin{document}


\title{Ion-lithium collision dynamics studied with a laser-cooled in-ring target}


\author{D. Fischer}
\email[]{fischer@mpi-hd.mpg.de}
\affiliation{Max Planck Institute for Nuclear Physics, Saupfercheckweg 1, 69117 Heidelberg, Germany}

\author{D. Globig}
\affiliation{Max Planck Institute for Nuclear Physics, Saupfercheckweg 1, 69117 Heidelberg, Germany}

\author{J. Goullon}
\affiliation{Max Planck Institute for Nuclear Physics, Saupfercheckweg 1, 69117 Heidelberg, Germany}

\author{M. Grieser}
\affiliation{Max Planck Institute for Nuclear Physics, Saupfercheckweg 1, 69117 Heidelberg, Germany}

\author{R. Hubele}
\affiliation{Max Planck Institute for Nuclear Physics, Saupfercheckweg 1, 69117 Heidelberg, Germany}

\author{V.L.B. de Jesus}
\affiliation{Instituto Federal de Educa\c{c}\~ao, Ci\^encia e Tecnologia do Rio de Janeiro (IFRJ),  Rua Lucio Tavares 1045, 
 26530-060  Nil{\'o}polis/RJ, Brazil}

\author{A. Kelkar}
\altaffiliation{present address: Institut f\"ur Ionenphysik und Angewandte Physik, Universit\"at Innsbruck, A-6020 Innsbruck, Austria }
\affiliation{Max Planck Institute for Nuclear Physics, Saupfercheckweg 1, 69117 Heidelberg, Germany}
\affiliation{Extreme Matter Institute EMMI, GSI Helmholtzzentrum f\"ur Schwerionenforschung GmbH, Planckstrasse 1, 64291 Darmstadt, Germany}

\author{A. LaForge}
\affiliation{Max Planck Institute for Nuclear Physics, Saupfercheckweg 1, 69117 Heidelberg, Germany}

\author{H. Lindenblatt}
\affiliation{Max Planck Institute for Nuclear Physics, Saupfercheckweg 1, 69117 Heidelberg, Germany}

\author{D. Misra}
\altaffiliation{present address: Tata Institute of Fundamental Research, Colaba, Mumbai 400 005, India}
\affiliation{Max Planck Institute for Nuclear Physics, Saupfercheckweg 1, 69117 Heidelberg, Germany}

\author{B. Najjari}
\affiliation{Max Planck Institute for Nuclear Physics, Saupfercheckweg 1, 69117 Heidelberg, Germany}

\author{K. Schneider}
\affiliation{Max Planck Institute for Nuclear Physics, Saupfercheckweg 1, 69117 Heidelberg, Germany}
\affiliation{Extreme Matter Institute EMMI, GSI Helmholtzzentrum f\"ur Schwerionenforschung GmbH, Planckstrasse 1, 64291 Darmstadt, Germany}

\author{M. Schulz}
\affiliation{Physics Department and LAMOR, Missouri University of Science \& Technology, Rolla, MO 65409, USA}
\affiliation{Institut f\"ur Kernphysik, Universit\"at Frankfurt, Max-von-Laue Strasse 1, 60438 Frankfurt, Germany}

\author{M. Sell}
\affiliation{Max Planck Institute for Nuclear Physics, Saupfercheckweg 1, 69117 Heidelberg, Germany}


\author{X. Wang}
\affiliation{Max Planck Institute for Nuclear Physics, Saupfercheckweg 1, 69117 Heidelberg, Germany}
\affiliation{Shanghai EBIT Laboratory, Institute of Modern Physics, Fudan University, Shanghai 200433, China}



\date{\today}

\begin{abstract}
We present a novel experimental tool allowing for kinematically complete studies of break-up processes of laser-cooled atoms. This apparatus, the 'MOTReMi', is a combination of a magneto-optical trap (MOT) and a Reaction Microscope (ReMi). Operated in an ion-storage ring, the new setup enables to study the dynamics in swift ion-atom collisions on an unprecedented level of precision and detail. In first experiments on collisions with 1.5\,MeV/amu O$^{8+}$-Li the pure ionization of the valence electron as well as ionization-excitation of the lithium target was investigated.
\end{abstract}


\maketitle


The description of the motion of few mutually interacting particles is one of the most fundamental and, at the same time, challenging tasks in physics. 
Break-up processes of atomic and molecular systems due to charged particle impact or photon absorption provide a very well suited testing ground for studying the dynamics of such correlated few-particle systems. Even though the underlying force, the electro-magnetic interaction, is well understood, the solution to the equations of motion is by no means trivial. Only recently it became possible to accurately predict differential cross sections of many basic systems, such as electron impact ionization of atomic hydrogen \cite{Rescigno99,PhysRevLett.89.273201} or photo double ionization of helium \cite{PhysRevLett.89.273201,PhysRevA.77.043420,PhysRevA.77.043421}. For these processes, impressive agreement to experimental data has been achieved.

For ion-impact, however, the situation is in many respects more complex.  Even though significant progress has been made \cite{Foster04,PhysRevA.81.042704,Colgan11}, the theoretical tools to describe ion-atom collisions are not as successful as for electron or photon impact 
\cite{Schulz03}. On the other hand, ions are in many respects a more versatile projectile species than electrons or photons. Interaction strengths can be varied from a photon-like perturbative regime to a strongly non-perturbative region at extremely small velocities which for electrons would bring the collision energy below the ionization threshold. On the other hand, in relativistic collisions with highly charged projectiles the target particles are exposed to the shortest (zeptosecond) and most intense electro-magnetic pulses that to date are accessible in laboratories. Thus, ion collisions provide also benchmarks for theoretical models in very 'exotic' situations.

From an experimental perspective the investigation of ion-atom collision dynamics represents a major challenge, too. Due to the large mass of the projectiles, their relative momentum change in the collision is often immeasurably small. As a result, for swift ionizing collisions, kinematically complete experiments only became feasible with the development of 'Reaction Microscopes' (or COLTRIMS) \cite{Doerner00,Ullrich03}. In this approach, the momentum vectors of recoiling target ions and electrons are measured directly and the momentum change of the projectile ion is obtained via momentum conservation. For almost two decades, this technique has been extensively applied to investigate e.g.\ single \cite{Schulz03,Fischer03,PhysRevLett.94.243201}, double \cite{Fischer03a,PhysRevA.80.062703}, and triple ionization \cite{PhysRevA.61.022703}, mutual ionization of the projectile and the target \cite{PhysRevLett.88.103202,PhysRevA.84.022707}, charge transfer \cite{PhysRevLett.76.3679,PhysRevA.85.022707}, or simultaneous transfer and ionization \cite{PhysRevLett.79.387,PhysRevLett.108.043202}. Even fully differential cross sections became accessible which represent the most stringent test of theoretical models. 

The accurate measurement of the collisionally induced recoil of the target ion requires thermal momenta that are relatively small. In most earlier experiments, rare gas atoms or molecular gases were used as targets, because these gases can efficiently be cooled to the required temperatures of about 1\,K or below, taking advantage of supersonic expansion in gas jet targets. In few measurements also atomic hydrogen was used \cite{PhysRevLett.103.053201}, though the dissociation of the molecules, e.g.\ by means of microwave fields, is intrinsically connected to a heating of the target which, in turn, limits the achievable resolution.

Laser-cooling in magneto-optical traps (MOT) has also been employed for the preparation of alkaline metal targets in several so-called MOTRIMS setups \cite{PhysRevA.62.043408,PhysRevLett.87.123201,PhysRevLett.87.123202,PhysRevLett.87.123203,blieck:103102,PhysRevLett.103.103008,DePaola2008139}. In these experiments the thermal momentum spread is typically in the range of 0.01\,a.u.\ resulting in an improved momentum resolution compared to most conventional Reaction Microscopes where the target temperature can be larger by up to 3 orders of magnitude. However, so far kinematically complete studies of ionizing collisions have not been possible as inhomogeneous magnetic fields, required for the trapping in the MOT, hampered the momentum resolved electron detection \cite{PhysRevA.83.023413}. All attempts to overcome this difficulty failed due to slowly decaying eddy currents in the vicinity of the electron trajectories. To date, there is no fully-functional Reaction Microscope equipped with a MOT target reported in the literature.

In this letter we report on the first realization of a MOTReMi, i.e.\ a Reaction Microscope with a magneto-optically trapped target. In the present case lithium is used as target which recently gained theoretical interest \cite{PhysRevLett.100.063002,PhysRevLett.108.053001}, because it represents the next step in complexity after helium. Moreover, lithium is particularly interesting for its -- in terms of electronic correlation -- asymmetric structure with only one valence electron and two strongly correlated K-shell electrons. In first experiments performed at the ion storage ring TSR in Heidelberg, the achieved momentum resolution was significantly improved compared to earlier studies with gas jet targets (e.g.\ \cite{PhysRevA.75.062708}). The electron cooling technique employed in the TSR allows for coherent projectile beams which recently were proven to be of crucial importance for the comparison of experimental data to quantum-mechanical models \cite{PhysRevLett.106.153202,Wang12}. Thus, the combination of these three techniques -- Reaction Microscope, MOT, and ion storage ring -- represents an acutely powerful experimental tool for the study of few-particle Coulomb-dynamics in fast ion-atom collisions.

\begin{figure}
\includegraphics[width=3.4in]{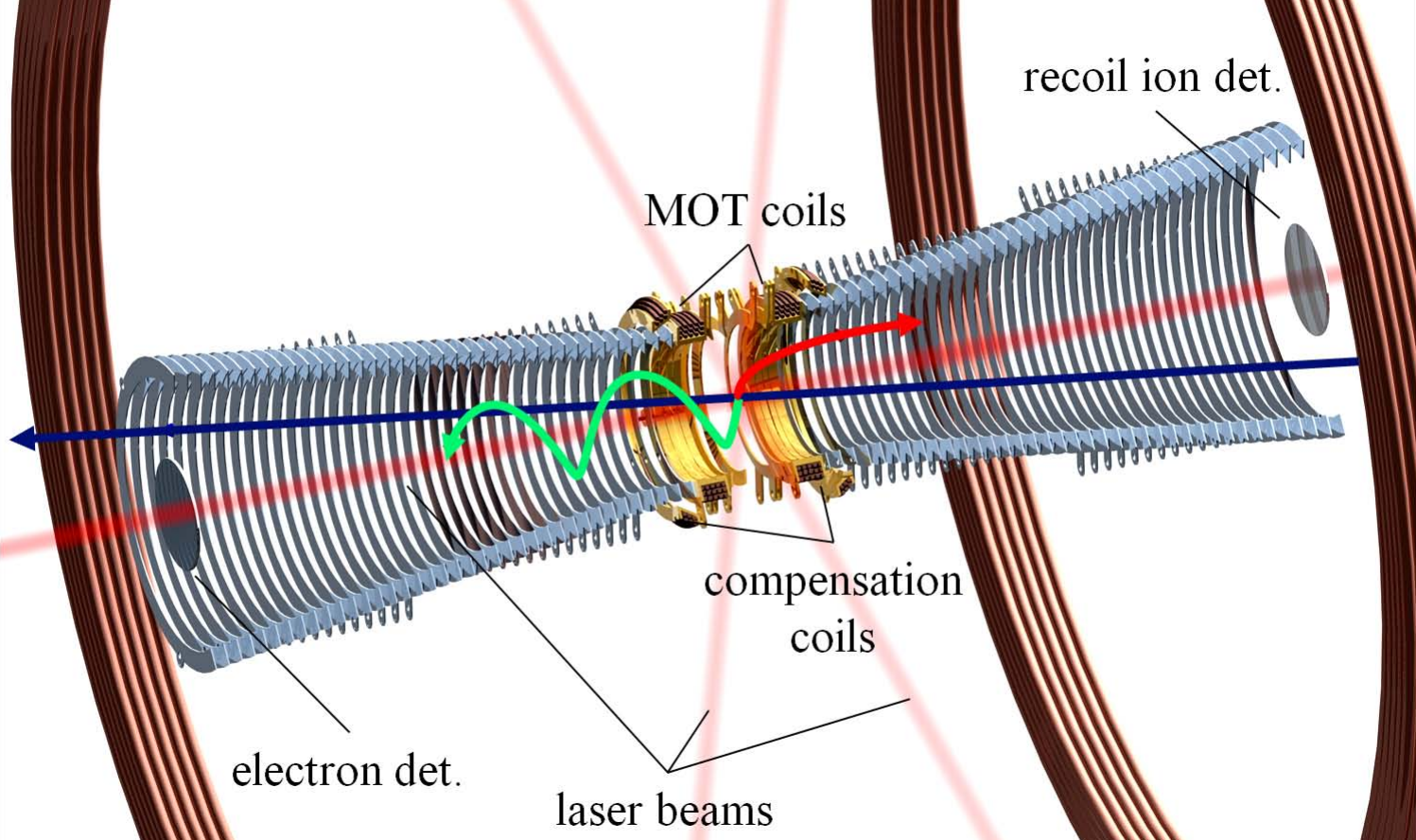}%
\caption{(Color online)  The MOTReMi apparatus.\label{fig_ MOTReMi}}
\end{figure}

The design of the  MOTReMi is illustrated in Fig.~\ref{fig_ MOTReMi}. Particular attention has been paid to two requirements that are indispensable for the optimal operation of these seemingly incompatible technologies: First, the MOT magnetic coils are kept as small as possible. These coils generate magnetic field gradients at the target position that are needed to trap the target atoms. A small spatial extension of the magnetic field allows for a fast switching which is, as will be detailed below, a crucial prerequisite for the momentum resolved detection of electrons. Second, a large opening for the projectiles of up to 100\,mm is required due to large projectile beam sizes and variations in the beam position directly after the ion injection in the storage ring. Not until 1\,s after the injection, the projectile beam size will shrink due to electron cooling to a diameter of roughly 1\,mm.

The electric extraction field is generated by means of ring electrodes which are distributed over a length of 84\,cm symmetrically to the reaction region. The spectrometer is vertically inclined by 8$^\circ$ with respect to the projectile beam direction and it tapers towards its center in order to simultaneously provide a large aperture for the ion beam and permit small MOT coils close to the target position. In contrast to all earlier MOT momentum spectrometers \cite{PhysRevA.62.043408,PhysRevLett.87.123201,PhysRevLett.87.123202,PhysRevLett.87.123203,blieck:103102,PhysRevLett.103.103008,DePaola2008139}, in the present setup a coaxial configuration of the MOT coils and the spectrometer is chosen which allows for a smaller coil diameter. The position sensitive particle detectors are centered with respect to the spectrometer axis. In order to increase the opening for the ion beam during injection, both detectors can be moved away from the beam axis by means of manipulators. 

The cooling and trapping of the target requires three orthogonal pairs of counter-propagating laser beams, two of which are directed perpendicular to the spectrometer axis through a gap between the two innermost ring electrodes. In conventional MOTs the third pair of laser beams is oriented coaxial to the coils, however, in the present setup this direction is blocked by the particle detectors. Here, the laser beams are tilted by 12$^\circ$.

The main difficulty of the adoption of laser cooling in Reaction Microscopes is the rapid switch-off of the MOT magnetic field enabling a momentum resolved electron detection. the fluctuations of the magnetic field should not exceed about 10\,mG. Because at zero field the cloud of trapped atoms expands very rapidly, the actual measurement should take place within a few milliseconds after the switch-off in order to maintain sufficient target density. This means, magnetic field decay times well below 1\,ms are required.

Such a switching performance is difficult to achieve, because induced fields generated by eddy currents decay on time scales of several milliseconds. To overcome these difficulties, several techniques have been developed, taking advantage of compensating current wave forms \cite{Dedman01} or oscillating magnetic fields in a so-called 'ac MOT' \cite{PhysRevLett.101.173201}. In the present setup we avoid eddy currents e.g.\ in the walls of the vacuum chamber by keeping the MOT magnetic field spatially as confined as possible. This is realized by keeping the overall size of the coils small and by employing a second pair of anti-Helmholtz coils which is slightly larger than the MOT coils and with an opposing current. These compensation coils efficiently reduce the range of the magnetic field without affecting the trapping efficiency, which relies on the magnetic field gradients in a rather small region around the trap position. Compared to an earlier setup \cite{PhysRevA.83.023413}, in the present case the magnetic field strength at the walls of the vacuum chamber are reduced by two orders of magnitude keeping the same field gradients in the center of the MOT.

Another complication for the operation of the MOT arises from the use of a homogeneous magnetic field in the Reaction Microscope. This field is oriented parallel to the spectrometer axis and it increases the transverse momentum acceptance for the electrons. In conventional MOTs the atoms are cooled with circularly polarized light in a $\sigma^+$-$\sigma^-$ configuration and are trapped at the position of the zero crossing of the magnetic field. The superposition of the Reaction Microscope's magnetic field with the MOT field causes a displacement of the point of zero field resulting in a shift of the target position along the spectrometer axis. Moreover, during the decay time of the MOT field, this shift even increases. It is therefore problematic to keep the cooling beams on and maintain optical molasses during magnetic field changes. In the present setup we avoid this problem by using two counterpropagating laser beams with the same polarization ($\sigma^+$-$\sigma^+$ or linearly polarized light) for cooling the target along the spectrometer axis. In this configuration, trapping still takes place (e.g.\ \cite{Hoepe94}) though the trapping position along this axis is largely independent on the magnetic field.

The current of the MOT coils is switched by means of MOSFETs. In standard operation, the field is enabled for 4\,ms per cycle with gradients of about 10\,G/cm and 5\,G/cm in the axial and radial directions, respectively. Leaving the current switched off for 2\,ms a recapture efficiency of close to 100\,\% is achieved, i.e.\ the MOT features a lifetime of several seconds even without reloading from our 2D-MOT beam source (similar to the one described in \cite{PhysRevA.80.013409}). About 250$\mu$s after the switch-off, no significant effect of stray magnetic fields on the electron momentum resolution is observed. The actual measurement takes place during a period of 1700\,$\mu$s/cycle. This corresponds to an overall duty-cycle of about 25\,\%. Optionally, the cooling laser beams can be turned off during the measurement period in order to provide 100\,\% ground state population of the target.

The  MOTReMi was commissioned using a pulsed 266\,nm Microchip laser. At this wavelength ($E_\gamma=4.65$\,eV), lithium can be ionized with an excess energy of 1.1\,eV from the excited 2$^2$P$_{3/2}$ state which is populated in the cooling transition. This allows one to accurately determine the resolution and momentum calibration. For the standard configuration with symmetric spectrometer potentials between 9\,V and -9\,V  we obtained a momentum resolution along the extraction direction of 0.02 and 0.06\,a.u.\ (FWHM) for electrons and recoil ions, respectively. In the perpendicular direction the resolution was about 0.1\,a.u.\ for both. This recoil ion momentum resolution is by a factor of 2 to 3 better than for comparable experiments with gas jet targets (e.g.\ \cite{Schulz03, PhysRevA.75.062708}) and better than in essentially all earlier MOTRIMS experiments \cite{blieck:103102}. For weaker electric fields, even better resolutions were attainable, however, at the expense of smaller electron energy acceptance. The magnetic field strength of the Reaction Microscope was about 8\,G resulting in a transverse electron energy acceptance of 20\,eV.

\begin{figure}
\includegraphics[width=2.6in]{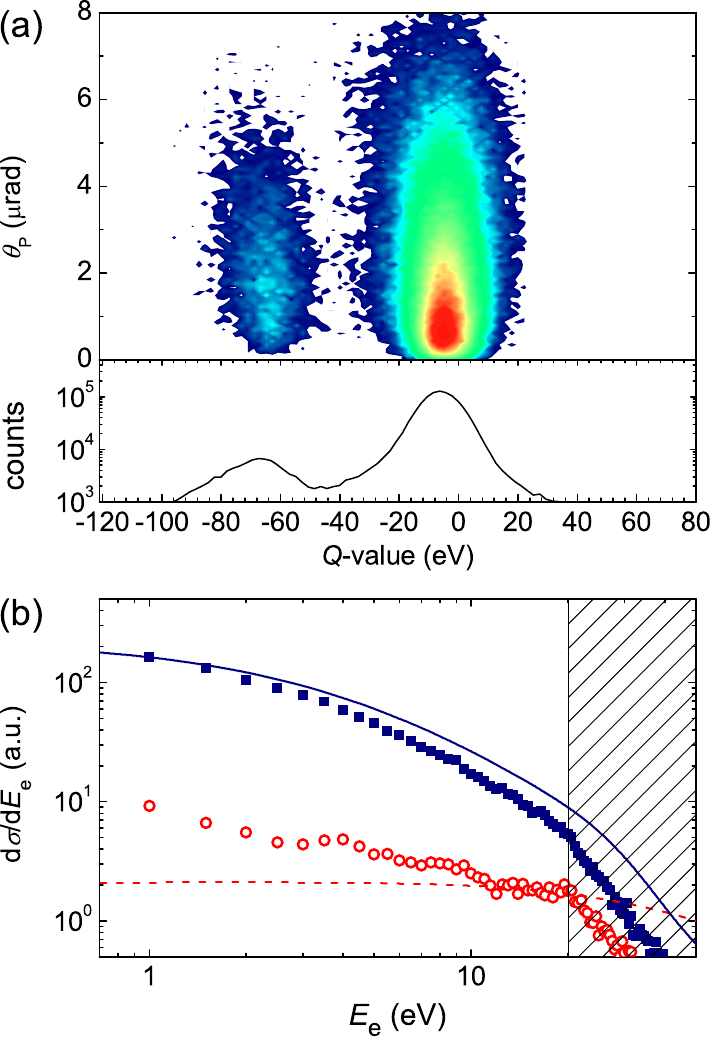}%
\caption{(Color online)  (a) Cross sections of single ionization in 1.5\,Mev/u O$^{8+}$-Li collisions as a function of the $Q$-value and the projectile scattering angle $\theta_{\text P}$. (b) ${\text d}\sigma /{\text d}E_{\text e}$ of the same collision system for pure 2s ionization (solid squares) and ionization-excitation (open circles). The curves are CDW-EIS results for 2s (solid line) and 1s (dashed line) ionization. In the shaded range a significant fraction of the electrons is not detected due to acceptance limitations.\label{fig2}}
\end{figure}
 
In the maiden experiment, single ionization of lithium in collisions with 1.5\,MeV/u O$^{8+}$ was investigated in the ion storage ring TSR. The $Q$-value of each collision event, i.e.\ its inelasticity, was calculated by $Q=E_{\text e}-q_\parallel \cdot v_{\text P}$  exploiting energy and momentum conservation ($q_\parallel$: longitudinal momentum transfer,  $v_{\text P}$: projectile velocity, $E_{\text e}$: final electron kinetic energy). In Fig.~\ref{fig2}\,(a) the cross section is plotted as a function of the $Q$-value and the projectile scattering angle $\theta_{\text P}$ which is not measured directly, but is calculated through $\theta_{\text P}\approx q_\perp/p_{\text P}$ (for $q/p_{\text P} \ll 1$, $q_{(\perp )}$: (transverse) momentum transfer, $p_{\text P}$: initial projectile momentum). The width of the angular projectile distribution seen in the figure does not reflect the experimental resolution (which is as small as 0.5\,$\mu$rad), but it is due to true physics effects. The recorded $Q$-value spectrum features two well-separated peaks. The dominant one centers at the ionization potential of the 2s state ($-5.4$\,eV) and corresponds to the ejection of L-shell electrons. The second one is shifted to about $-65$\,eV and represents ionization events where the Li$^+$ ion remains finally in an excited state $n\,l$ ($n\geq 2$). 

The single differential cross sections ${\text d}\sigma /{\text d}E_{\text e}$ of the two reaction channels are plotted in Fig.~\ref{fig2}\,(b) and compared to a CDW-EIS model using one-electron Hartree-Fock wave functions. The experimental data are normalized to the calculated 2s ionization cross section at zero electron energy. In both experimental distributions a sudden change of slope is observed at $E_{\text e}=20$\,eV which is due to the limited acceptance of the spectrometer in the direction perpendicular to the spectrometer axis. In principle, the acceptance for electrons emitted in the backward direction is restricted to even lower energies (with the chosen spectrometer voltages the limit is at 9\,eV). However, for such large perturbations as in the present collision system ($Z_{\text P}/v_{\text P}=1.0$\,a.u.) the electrons are focused in the forward direction due to the post collision interaction, and already at lower energies the contribution of backwards ejected electrons becomes insignificant.

For L-shell ionization, the measured cross section drops slightly faster than the theoretical curve, but overall there is fair agreement. In the second reaction channel, the excitation of the  Li$^+$ recoil ions can occur due to several mechanisms. The simplest one corresponds to the emission of a K-shell electron through the interaction with the projectile while the valence electron remains unaffected as a spectator \cite{PhysRevLett.80.4649}. This process is calculated with our model and shown in the graph. Here, only poor agreement between theory and experiment is achieved,  in shape as well as in relative magnitude. In particular for electron energies below 10\,eV the model considerably underestimates the cross section.

In the effective one-electron model used above, some important aspects are neglected. First, due to the large binding energy of the K-shell electron, the ionization process takes place only for relatively close collisions. It seems likely, that in such close collisions the valence electron, whose ionization potential is more than ten times smaller, is ejected, too. This means, that the ionization of the 1s electron will often result in double ionization which, in turn, would even increase the observed discrepancies. In a second two-electron mechanism not considered in the model, the inner-shell electron is promoted to an excited state in the target. This excitation is also expected to occur in close collisions, which again likely will result in the emission of the valence electron. Thereby, the inner-shell excitation will effectively contribute to single ionization and explain the observed discrepancies. This gives an indication that the simultaneous excitation and ionization is a prominent and for electron energies below 10\,eV even dominant process resulting in the core hole creation in ionization processes of the lithium target.

The simultaneous excitation and ionization in ion-lithium collisions \cite{PhysRevLett.83.1131} has been observed earlier using Auger-spectroscopy. This process has also been the subject of many earlier experimental \cite{PhysRevLett.95.033201,PhysRevLett.96.223201} as well as theoretical investigations \cite{PhysRevA.81.042712,PhysRevLett.107.023203} for electron collisions on a helium target. In all these studies, ionization-excitation was mainly considered to be a sensitive probe for the correlation between the two active electrons in the K-shell. In the present case the situation is rather different. On the one hand, in the present experiment the two active electrons are initially in different shells which means that their interaction is much weaker than for the two ground state electrons. On the other hand, due to the large perturbation in the present collision system, ionization plus excitation is much more likely to proceed through a higher-order process involving at least two independent interactions with the projectile. 

The important role of this independent higher-order channel leads to some interesting insight to be gained from the data. First, studying the electron ejection from the valence shell, the excitation process restricts the collision to small impact parameters, but should otherwise not affect the electron emission characteristics. In other words, the electron emission pattern of this process corresponds to an impact-parameter selective cross section for the 2s ionization. Second, concentrating on the excitation process, the ejection of the loosely-bound electron only serves as 'marker' for the target atom to be able to use the COLTRIMS technique that is not applicable for neutral particles. This way, differential data on target excitation becomes accessible, e.g.\ in the projectile scattering angle, which can be measured by other means only for much smaller collision energies and with lower resolution \cite{PhysRevLett.40.1646}.

In conclusion we have reported on the first successful operation of a  MOTReMi. The inherent difficulties related to the magnetic field switching have been solved by using special configurations of the MOT magnetic field as well as the cooling laser polarizations. As a result, a new experimental tool is provided which enables the kinematically complete study of atomic fragmentation processes in essentially all conceivable kinds of collisions involving ions, electrons, photons, or strong fields from pulsed lasers or FEL facilities. Compared to conventional Reaction Microscopes, the new technique offers significantly higher resolutions due to much lower target temperatures. Moreover, it makes alkaline metal atoms accessible for collision experiments which are interesting for the simple structure having one optically active electron.

In the inaugural experiment of the MOTReMi, single ionization as well as ionization-excitation in 1.5\,MeV O$^{8+}$-lithium was studied. Due to the weak initial state correlation between the inner and outer shell electrons, the latter process provides new insights into the dynamics of both the electron ejection process as well as the collisionally induced excitation of the atomic target.  

\begin{acknowledgments}
We would like to thank S.~Jochim, A.~Dorn, R.~Moshammer and their group members for many fruitful discussions and useful hints. We are grateful for the steady encouragement of J.~Ullrich. Furthermore, we acknowledge the experimental support of C. Krantz and the excellent work of the MPIK accelerator and TSR teams. This work was funded by the German Research Council (DFG), under grant No.\ FI 1593/1-1. We are grateful for the support by the Alliance Program of the Helmholtz Association (HA216/EMMI). M.S.\ acknowledges the support by the National Science Foundation, under grant No.\ 0969299,  by the DFG, and by the Fulbright Foundation. V.L.B.J. is grateful for the support by DAAD, CAPES, and IFRJ.
\end{acknowledgments}



%

\end{document}